\documentclass{aa}
\usepackage{graphicx}
\begin{document}

\title{\bf Curious Variables Experiment (CURVE).\\
CCD Photometry of QW Serpentis in Superoutburst and Quiescence}
\author{A. Olech\inst{1},
P. K\c{e}dzierski\inst{2}, K. Z{\l}oczewski\inst{2},
K. Mularczyk\inst{2}  and M. Wi{\'s}niewski\inst{1}}

\offprints{A. Olech}

   \institute{Nicolaus Copernicus Astronomical Center, ul. Bartycka 18,
        00-716 Warszawa, Poland \\
        \email{olech@camk.edu.pl}
        \and
    Warsaw University Observatory, Al. Ujazdowskie 4, 00-478 Warszawa, Poland\\
        \email{(pkedzier,kzlocz,kmularcz,mwisniew)@astrouw.edu.pl}}

   \date{Received .................., 2003; accepted ................ 2003}

   \abstract{
We report extensive photometry of the dwarf nova QW Ser throughout its
2003 February superoutburst till quiescence. During the superoutburst the star
displayed clear superhumps with a mean period of $P_{sh} = 0.07703(4)$
days. In the quiescence we observed a double humped wave characterized
by a period of $P=0.07457(2)$ days. As both periods obey the Stolz-Schoembs
relation with a period excess equal to $3.30\pm0.06$\% the latter period
is interpreted as the orbital period of the binary system.
}

\titlerunning{CURVE. CCD Photometry of QW Serpentis in Superoutburst and Quiescence}
\authorrunning{Olech et al.}
   \maketitle
%
%________________________________________________________________

\section{Introduction}

QW Ser (TmzV46, USNO-A2.0 0975-07829422) was discovered by Takamizawa
(1998), who detected three significant brightenings of the star on JD
2449620, 2449716 and 2450906. Photometric behavior and color of QW Ser
suggested that it was a dwarf nova. In October 1999, Schmeer (1999)
reported another outburst of QW Ser. Subsequent observations were made
by Kato \& Uemura (1999). According to them, the initial decline of
$0.73\pm0.20$ mag observed during the first seven days of the outburst,
was followed by a much slower decrease of the brightness. The reported
rate of mean decline of 0.1 mag per day was neither consistent with
those of superoutburst in SU UMa-type stars nor long outbursts of SS
Cyg-type dwarf novae, thus the definitive classification of QW Ser
remained an open question.

The 2003 February outburst of QW Ser was reported by Muyllaert (2003),
who caught the star at a magnitude of 12.6 on 2003 Feb. 23.715 UT.
Reported at a later date observations of Schmeer (2003) indicated that
star reached 12.5 mag on Feb. 21.199 UT. 

\section{Observations and Data Reduction}

Observations of QW Ser reported in the present paper were obtained during
14 nights between February 23, 2003 and May 04, 2003 at the
Ostrowik station of the Warsaw University Observatory. The data was
collected using the 60-cm Cassegrain telescope equipped with a
Tektronics TK512CB back illuminated CCD camera. The scale of the camera
was 0.76"/pixel providing a $6.5'\times 6.5'$ field of view. The full
description of the telescope and camera was given by Udalski and Pych
(1992).

We monitored the star in ``white light''. This was due to the lack of
an autoguiding system, not yet implemented after a recent telescope
renovation. Thus we did not use any filter to shorten the exposures in
order to minimize guiding errors. 

The exposure times were from 60 to 90 seconds during the bright state
and from 200 to 350 seconds in the minimum light.

A full journal of our CCD observations of QW Ser is given in Table
1. In total, we monitored the star during 36.47 hours and obtained 896
exposures.

\begin{table}[h]
\caption{Journal of the CCD observations of QW Ser}
\begin{tabular}{|l|c|c|r|r|}
\hline
\hline
Date of& Start & End & Length & No. of \\
2003   & 2452000. + & 2452000. + & [hr]~ & frames \\
\hline
Feb. 23/24 &694.53928 &694.68717 & 3.549 & 174\\
Feb. 24/25 &695.50857 &695.69560 & 4.489 & 197\\
Feb. 25/26 &696.55102 &696.68165 & 3.135 & 160\\
Feb. 26/27 &697.55928 &697.68020 & 2.902 & 136\\
Mar. 07/08 &706.61633 &706.66486 & 1.165 &  17\\
Mar. 24/25 &723.49079 &723.60672 & 2.782 &  28\\
Mar. 26/27 &725.46410 &725.58953 & 3.010 &  30\\
Mar. 29/30 &728.47971 &728.60005 & 2.888 &  24\\
Mar. 31/01 &730.48924 &730.59647 & 2.574 &  23\\
Apr. 24/25 &754.46824 &754.58584 & 2.822 &  29\\
Apr. 25/26 &755.43301 &755.50438 & 1.713 &  21\\
Apr. 29/30 &759.46412 &759.47453 & 0.250 &   3\\
May  03/04 &763.43619 &763.53926 & 2.473 &  23\\
May  04/05 &764.41933 &764.53249 & 2.716 &  31\\
\hline
Total          &   --   & -- & 36.468 & 896 \\ 
\hline
\hline
\end{tabular}
\end{table}

All the data reductions were performed using a standard procedure
based on the IRAF \footnote{IRAF is distributed by the National Optical
Astronomy Observatory, which is operated by the Association of
Universities for Research in Astronomy, Inc., under a cooperative
agreement with the National Science Foundation.} package and
the profile photometry has been derived using the DAOphotII package
(Stetson 1987).

Relative unfiltered magnitudes of QW Ser were determined as the
difference between the magnitude of the variable and the magnitude of
the comparison star GSC 0927:464 ($R.A.=15^h26^m18.1^s$, $Decl. =
+08^\circ17'50.3"$) located 1.1 arcmin to the east of the variable. This
comparison star is marked in the chart displayed in Fig. 1. 

   \begin{figure}[!h]
   \centering
\includegraphics[scale=.49]{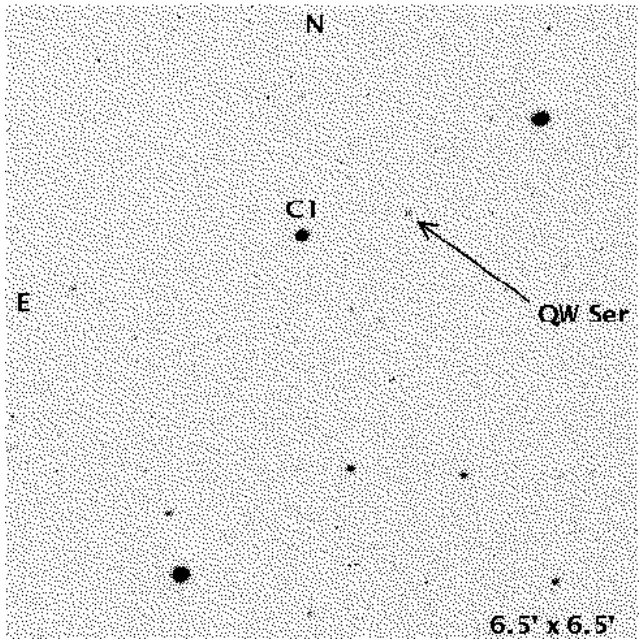}
      \caption{Finding chart for QW Ser covering a region of $6.5 \times
6.5$ arcminutes. The position of the comparison star is shown. North is
up, East is left.
              }
         \label{Fig1}
   \end{figure}

   \begin{figure}
   \centering
\includegraphics[bb=90 230 600 520,scale=0.59]{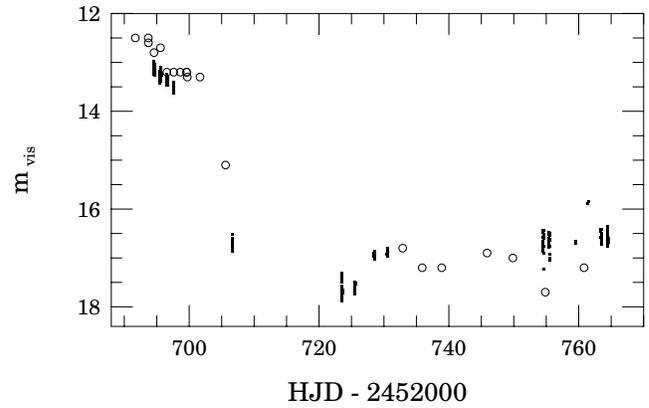}
      \caption{ The general photometric behavior of QW Ser during
its 2003 superoutburst. Visual estimates collected in the VSNET archive
are shown as open circles and CCD observations described in this
paper as dots.
              }
         \label{Fig2}
   \end{figure}

   \begin{figure}
   \centering
\includegraphics[bb=85 245 520 750,scale=0.64]{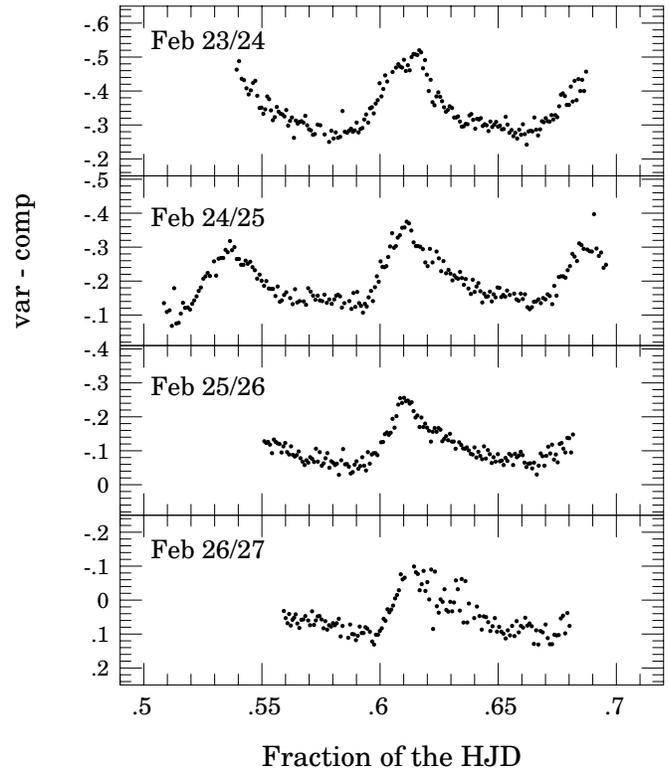}
      \caption{The light curves of QW Ser observed during four
consecutive nights in February 2003.
              }
         \label{Fig3}
   \end{figure}

The typical accuracy of our measurements varied between 0.001 and 0.009
mag in the bright state and between 0.009 and 0.091 mag in the minimum
light. The median value of the photometry errors was 0.004 and 0.031
mag, respectively.

\section{Light curves}

Fig. 2 presents the photometric behavior of QW Ser as observed between
February and May 2003. Relative magnitudes of the variable were
transformed to the visual scale using photographic magnitude of our
comparison star equal to $13.59\pm0.40$ mag and taken from GSC Catalog
(Lasker et al. 1999). Additionally, open circles denote the visual and
CCD estimates made by astronomy amateurs and published in the VSNET
mailing list. Our transformation of "white light" observations is based
on the photographic magnitude of the comparison star and we would like
to point our that it is a rough estimate, which is used only for showing
the general behavior of the star. The true brightness of the QW Ser may
differ even by about 0.5 magnitude from this one shown in Fig. 2.

It is difficult to determine the exact time of the beginning of the
superoutburst. The maximal brightness of 12.5 mag was reported by
Schmeer (2003) on Feb. 21.199 UT. Previous estimates of the brightness
of QW Ser were made a week earlier and found the star in quiescence.
Presence of the clear superhumps with an amplitude as high as 0.23 mag
during our first run on Feb. 23/24 may suggest that Schmeer's
observations correspond to a real beginning of the superoutburst. Thus
our first observing run is most probably the fourth night of the
superoutburst.

   \begin{figure}
   \centering
\includegraphics[bb=85 30 520 732,scale=0.64]{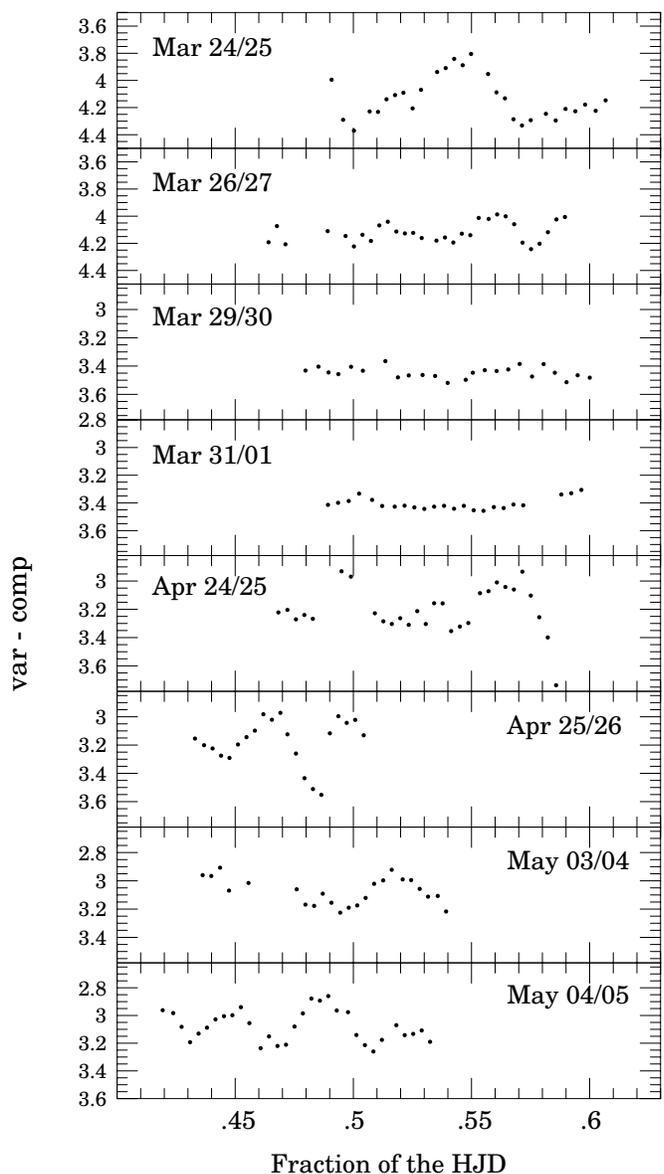}
      \caption{The light curves of QW Ser observed during eight
longest runs from period Mar. 07/08 -- May 04/05.
              }
         \label{Fig4}
   \end{figure}

   \begin{figure}
   \centering
\includegraphics[bb=65 250 500 620,scale=0.56]{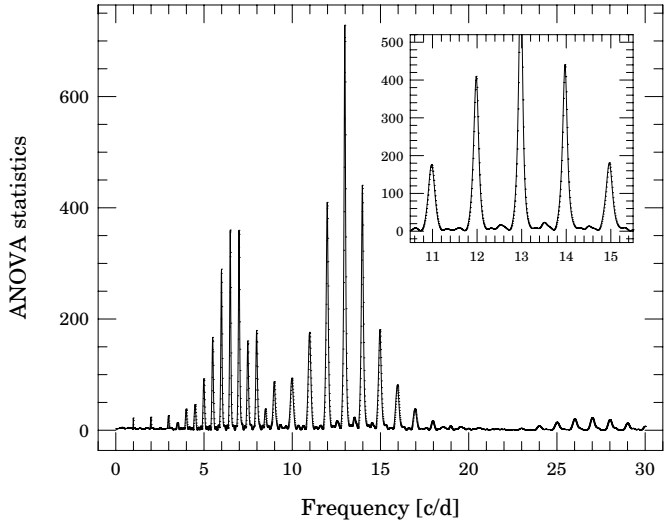}
      \caption{{\sc anova} power spectrum of the light curve of QW Ser.
The inset shows the magnification of the power spectrum around main
frequency.
              }
         \label{Fig5}
   \end{figure}

Our observations from Mar. 07/08 catch the star around minimum light.
Observations by Dubovsky (2003) made on Mar. 7.099 UT indicate that the
brightness of QW Ser was then at a magnitude of 15.1 i.e. about two
magnitudes brighter than our determination from subsequent night. Thus
we conclude that superoutburst of QW Ser lasted until Mar. 07/08 i.e. 15
days. 

Fig. 3 shows the light curves of QW Ser during four consecutive nights
of the 2003 February superoutburst. The superhumps are clearly visible
in each run. Their amplitude is 0.23, 0.21, 0.18 and 0.16 mag from the
first to fourth night, respectively.

The lights curves of the eight longest runs from the period Mar. 07/08
-- May 04/05 are displayed in Fig. 4. The magnitude of the star varied
during this interval with an amplitude sometimes exceeding 0.5 mag, thus
suggesting the presence of the periodic signal in the data (see Sec. 5).

\section{Superhumps}

From each light curve of QW Ser in superoutburst we removed the first or
second order polynomial and then analyzed them using {\sc anova}
statistics and two harmonics Fourier series (Schwarzenberg-Czerny 1996).
The resulting periodogram is shown in Fig. 5. The most prominent peak is
found at a frequency of $f=12.98\pm0.05$ c/d, which corresponds to the
period $P_{sh}=0.07704(30)$ days ($110.9\pm0.4$ min). The peak visible
at 6.5 c/d is a ghost of the main frequency arising due to the use of
two harmonics. The harmonic peak at 26 c/d appears to be real. The inset
in Fig. 5 shows the magnification of the power spectrum around the main
frequency. Apart from this main peak and its aliases the inset shows no
other significant periodicities.

The light curves of QW Ser in superoutburst were prewhitened with the main
period and its first harmonic. The power spectrum of the resulting light
curve shows no clear peaks except second, third and fourth harmonics of
the main frequency.

For nights from the superoutburst we determined six times of maxima and
nine times of minima of the superhumps. They are shown in Table 2
together with their cycle numbers $E$. The least squares linear fit to
the data from Table 2 gives the following ephemeris for the maxima:

\begin{equation}
{\rm HJD}_{max} =  2452694.6154(24) + 0.07674(13) \cdot E
\end{equation}

\noindent and for the minima:

\begin{equation}
{\rm HJD}_{min} =  2452694.5829(27) + 0.07726(11) \cdot E
\end{equation}

\begin{table}[h]
\caption{Times of extrema in the light curve of QW Ser in February 2003}
\begin{center}
\begin{tabular}{|l|c|r||l|c|r|}
\hline
\hline
\multicolumn{3}{|c||}{\bf Maxima} & \multicolumn{3}{|c|}{\bf Minima} \\
\hline
$E$ & HJD & O -- C & $E$ & HJD & O -- C\\
\hline
\hline
0 & 694.617 & 0.021 & 0 & 694.580 & $-0.037$ \\
12 & 695.536 & 0.003 & 1 & 694.661 & 0.011 \\
13 & 695.612 & $-0.019$ & 12 & 695.513 & 0.039 \\
14 & 695.685 & $-0.062$ & 13 & 695.593 & 0.074 \\
26 & 696.610 & $-0.008$ & 14 & 695.663 & $-0.020$ \\
39 & 697.613 & 0.062 & 26 & 696.589 & $-0.034$ \\
   &          &          & 27 & 696.665 & $-0.051$ \\
   &          &          & 39 & 697.598 & $0.019$ \\
   &          &          & 40 & 697.671 & $-0.030$ \\
\hline
\hline
\end{tabular}
\end{center}
\end{table}

Combination of both O -- C determinations of the period with its value
derived from the power spectrum analysis  gives the mean value of
superhump period as equal to $P_{sh} = 0.07703(4)$ days i.e.
$110.92\pm0.06$ min.

The $O - C$ departures from the ephemeris (1) and (2) are given also in
Table 2. They show no clear trend, thus we conclude that our
observational coverage of the superoutburst was too weak for a valuable
determination of the superhump period derivative.

\section{Quiescence}

During the period Mar. 07/08 -- May 04/05 QW Ser was observed in a
minimum light of around 17 mag. The {\sc anova} power spectrum for this
period is shown in Fig. 6. Before the calculation the light curves were
prewhitened using the first order polynomial.

   \begin{figure}
   \centering
\includegraphics[bb=65 250 500 620,scale=0.56]{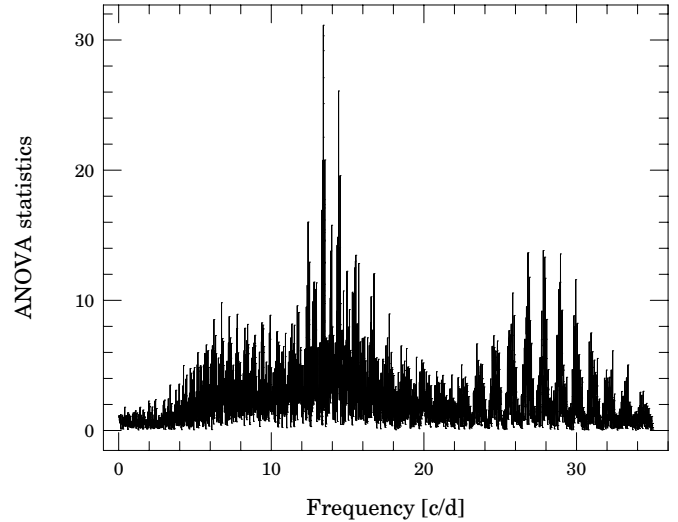}
      \caption{{\sc anova} power spectrum of the light curve of QW Ser
from period Mar. 07/08 -- May 04/05.
              }
         \label{Fig6}
   \end{figure}

The resulting periodogram yields a clear peak at a frequency of
$f=13.410\pm0.003$ c/d, which corresponds to a period of $P=0.07457(2)$
days ($107.38\pm0.03$ min).

A significant group of high peaks is also visible close to the first
harmonic of main period at a frequency around 26.8 c/d. It is due to the
shape of the light curve, which often displayed a double wave structure
(see Fig. 4 and nights Mar. 26/27, Apr. 25/26 and May 04/05, for
example).

We have prewhitenned our original light curve of QW Ser using the main
periodicity and its first harmonics. The power spectrum of the resulting
light curve shows no significant peaks in period range of 0--50 c/d
indicating that in quiescence QW Ser in a monoperiodic object.

\section{Discussion}

The first detection of the superhumps with a period of
$P_{sh}=0.07703(4)$ during the 2003 superoutburst of QW Ser unambiguously
proves that this star belongs to the group of SU UMa-type dwarf novae.

Superhumps occur at a period slightly longer than the orbital period of
the binary system. They are most probably a result of the accretion disc
precession caused by the gravitational perturbations from the secondary.
These perturbations are most effective when the disc particles moving on
the eccentric orbits enter the 3:1 resonance. Then the superhump period
is simply the beat period between orbital and precession rate periods:

\begin{equation}
\frac{1}{P_{sh}} = \frac{1}{P_{orb}} - \frac{1}{P_{prec}}
\end{equation}

The precession rate of the eccentric disc was first discussed by Osaki
(1985). Based on a nonresonant free-particle orbit at the disk edge he
derived the following expression for the precession rate:

\begin{equation}
\frac{P_{orb}}{P_{prec}} = \frac{3}{4}\frac{q}{\sqrt{1+q}}\left(\frac{R}{a}\right)^{3/2}
\end{equation}

\noindent where $a$ is the binary separation, $R$ is the disc radius and
$q$ is the mass ratio ${M_2}/{M_1}$. At the 3:1 resonance we can assume
that $R\approx0.46a$ and hence:

\begin{equation}
\frac{P_{orb}}{P_{prec}} \approx \frac{0.233q}{\sqrt{1+q}}
\end{equation}

Defining the period excess $\epsilon$ as:

\begin{equation}
\epsilon = \frac{\Delta P}{P_{orb}} = \frac{P_{sh} - P_{orb}}{P_{orb}}
\end{equation}

\noindent from equation (5) we can simply derive the relation between
the period excess and the mass ratio:

\begin{equation}
\epsilon\approx\frac{0.23q}{1+0.27q}
\end{equation}

From the observational point of view it was first noticed by Stolz \&
Schoembs (1984) that $\epsilon$ grows with $P_{orb}$. It fact this
relation is not only obeyed by the ordinary SU UMa stars but also
by the permanent superhumpers (Skillman \& Patterson 1993).

   \begin{figure}
   \centering
\includegraphics[bb=65 240 500 610,scale=0.55]{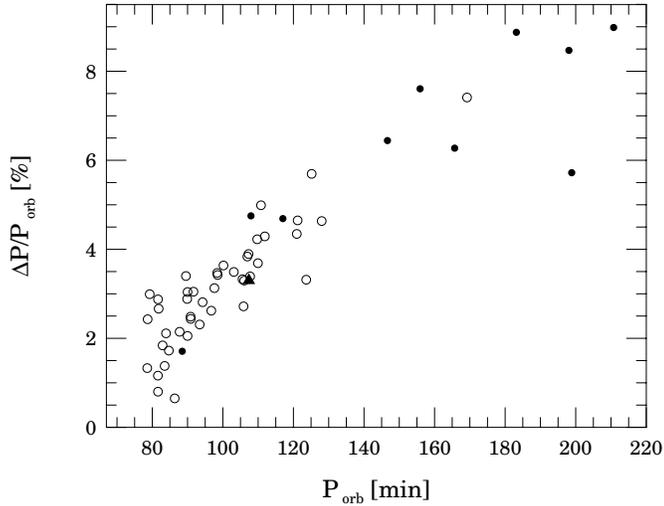}
      \caption{Stolz \& Schoembs relation for 44 ordinary SU UMa
stars (open circles) and 10 stars other than dwarf novae (filled
circles). The position of QW Ser is denoted by the solid triangle.
              }
         \label{Fig7}
   \end{figure}

Fig. 7 shows the Stolz \& Schoembs relation for 44 ordinary SU UMa-type
stars (open circles) and 10 stars other than dwarf novae (filled
circles). Data for this graph are taken from Patterson (1998) and
newest alerts published in the VSNET archive.

The double wave structure of brightness modulations observed in QW Ser
during quiescence and the period of these modulations which is slightly
shorter than the superhump period strongly suggest that we detected the
orbital period of the binary system. Double humped waves characterized
by the orbital period of the binary are observed in some SU UMa stars
with high inclination of the orbit (see for example WZ Sge, AL Com and
V485 Cen - Patterson et al. 1996, 2002, Olech 1997). The period excess
for QW Ser is then $\epsilon = 0.033\pm0.006$. According to the equation
(7) this corresponds to the mass ratio $q=0.15$.

In Fig. 7 QW Ser is plotted using a filled triangle. Error bars are of
the same size as the symbol. It is clear that both periods observed in
the light curve of this stars conform to the Stolz \& Schoembs
relationship. This is a strong argument for our interpretation that
modulations observed in quiescence are connected with the orbital period
of the binary system.

\begin{acknowledgements}
We gratefully acknowledge the generous allocation of  the Warsaw
Observatory 0.6-m telescope time. This work was partially supported by
KBN grant number 2~P03D~024~22 to A. Olech and used the on-line service
of the VSNET. We would like to thank Chris O'Connor for reading and
commenting on the manuscript.
\end{acknowledgements}

\end{document}